\documentclass[letterpaper, 10 pt, conference]{ieeeconf}  % Comment this line out if you need a4paper

\IEEEoverridecommandlockouts                              % This command is only needed if 
\usepackage{graphics} % for pdf, bitmapped graphics files
\usepackage{epsfig} % for postscript graphics files
\usepackage{amsmath} % assumes amsmath package installed
\usepackage{amssymb}  % assumes amsmath package installed
\usepackage[ruled,longend]{algorithm2e}
\usepackage{subfig}
\usepackage{balance}
\title{\LARGE \bf
T-PFC: A Trajectory-Optimized Perturbation Feedback Control Approach%Under Action Uncertainty
}

\author{Karthikeya S Parunandi$^{1}$ and Suman Chakravorty$^{2}$% <-this % stops a space
\thanks{$^{1}$Karthikeya S Parunandi is with the Department of Aerospace Engineering, Texas A\&M University, College Station
        {\tt\small s.parunandi@tamu.edu}}%
\thanks{$^{2}$Suman Chakravorty is with Faculty of Aerospace Engineering, Texas A\&M University, College Station
        {\tt\small schakrav@tamu.edu}}%
}

\begin{document}

\maketitle
\thispagestyle{empty}
\pagestyle{empty}

%%%%%%%%%%%%%%%%%%%%%%%%%%%%%%%%%%%%%%%%%%%%%%%%%%%%%%%%%%%%%%%%%%%%%%%%%%%%%%%%
\begin{abstract}
Traditional stochastic optimal control methods that attempt to obtain an optimal feedback policy for nonlinear systems are computationally intractable. In this paper, we derive a  decoupling principle between the open loop plan, and the closed loop feedback gains, that leads to a deterministic perturbation feedback control based solution (T-PFC) to fully observable stochastic optimal control problems, that is near-optimal to the third order. Extensive numerical simulations validate the theory, revealing a wide range of applicability, coping with medium levels of noise. The performance is compared with Nonlinear Model Predictive Control in several difficult robotic planning and control examples that show near identical performance to NMPC while requiring much lesser computational effort. %It also leads us to raise the bigger question as to why NMPC should be used in robotic control as opposed to perturbation feedback approaches. 
\end{abstract}

%%%%%%%%%%%%%%%%%%%%%%%%%%%%%%%%%%%%%%%%%%%%%%%%%%%%%%%%%%%%%%%%%%%%%%%%%%%%%%%%
\section{INTRODUCTION}

%- MDPs
%- iLQG,DDP
%- MPC
%- planning for nonholonomic robots
%- Contributions of this paper\\
Stochastic optimal control is concerned with obtaining control laws under uncertainty, minimizing a user-defined cost function while being compliant with its model and constraints. This problem frequently arises in robotics, where, planning a robot's motion under sensor, actuator and environmental limitations is vital to achieve a commanded task. At present, online planning methods such as Model Predictive Control (MPC) are preferred over offline methods. However, it takes a toll on the onboard computational resources. On the other hand, offline solutions are susceptible to drift, and cannot deal with a  dynamic environment. In this paper, we propose a composite approach that merges the merits of both approaches i.e, computation off-line and a robust feedback control online, while re-planning, unlike in MPC, is performed only rarely, and is typically required only beyond moderate levels of noise.

The main contributions of this paper are as follows:
(a) to demonstrate the decoupling between the deterministic open-loop and the closed loop feedback control of perturbations, in a fully-observed stochastic optimal setting, that is near-optimal to third order, (b) to propose a novel method based on the aforementioned decoupling principle to deal with Robotic stochastic optimal control problem, and (c) to draw  comparisons between the proposed approach and the non-linear MPC framework, aimed at re-examining the widespread use of non-linear MPC in Robotic planning and control.  

\begin{figure}[thpb]
      \centering
    \subfloat[Cost comparison]{{\includegraphics[width=0.5\linewidth]{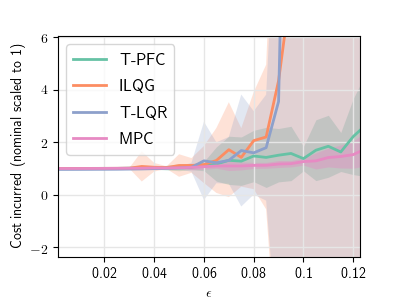}}}
    \subfloat[No. of re-plannings for $\epsilon > 0.25$]{{\includegraphics[width=0.5\linewidth, height=3.05cm]{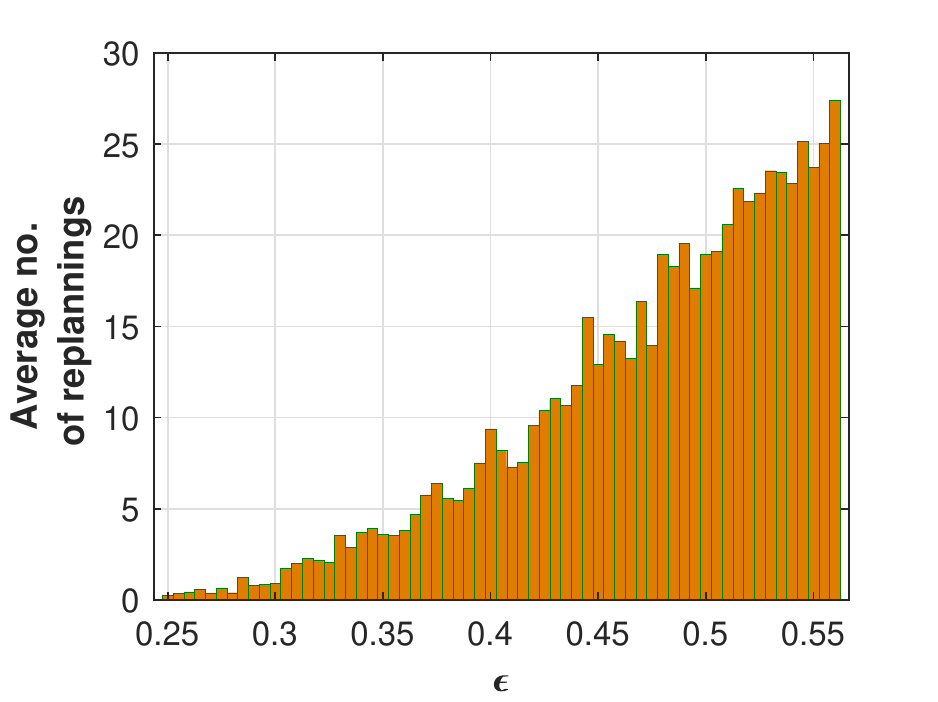} }}
    \caption{\small{(a) Cost evolution over a feasible range of $\epsilon$ for a car-like robot, where $\epsilon$ is a measure of the noise in the system. Note that the performance of T-PFC is close to NMPC for a wide range of noise levels, while T-PFC takes approximately  $100\times$ less time to execute (see Table I). (b) No. of re-plannings for above-moderate noise levels in the car-like robot simulation in gazebo using T-PFC is still around 8 times less than NMPC.}}%
    \label{cost_1}
   \end{figure}
   
\section{RELATED WORK}
In fully observable systems, decision-making is typically modeled as a Markov Decision Process (MDP). Methods that try to solve MDPs using dynamic programming/HJB face the `curse of dimensionality' in high-dimensional spaces while discretizing the state space \cite{c1}. Hence, most successful/practical methods are based on Pontryagin's maximum principle \cite{c2} though it results in locally optimal solutions. 
Iterative methods such as ILQG \cite{c3}, DDP \cite{c4} and stochastic DDP \cite{c5} fall under this category. They expand the optimal cost-to-go and the system dynamics about a nominal, which is updated with every iteration. 
ILQG relies on the quadratic expansion of the cost-to-go and a linear expansion of system dynamics. DDP/stochastic-DDP considers the quadratic approximation of both. The convergence of these methods is similar to Newton's method. These methods iteratively optimize the open loop and the linear feedback gain, in lieu, in our approach, owing to the decoupling, the open loop optimal control sequence is obtained using a state of the art nonlinear Programming (NLP) solver, and given this open loop sequence, the optimal feedback gain is obtained using the ``decoupled" gain equations. This, in turn, avoids, the expensive recursive Ricatti solutions required by ILQG and DDP techniques (also see Sec. IV).\\ %Though the variants of these methods are able to handle constraints, they are not as straightforward to deal with as compared to imposing them while solving a numerical optimization.\\ 

Model Predictive Control (MPC) is a popular planning and control framework in robotics. It bypasses the curse of dimensionality by repeatedly generating open-loop controls through the numerical solution of a finite horizon constrained optimal control problem at every discrete time-step \cite{c6}. Initially employed in chemical process industry \cite{c7}, MPC has found widespread application in Robotics owing to its ability to handle nonlinearity and constraints. Currently, this framework is well-established in the field and has demonstrated success in diverse range of problems including manipulation \cite{c9}, visual servoing \cite{c9}, and motion planning. In robotic motion planning, MPC is widely in use for motion planning of mobile robots, manipulators, humanoids and aerial robots such as quadrotors \cite{c11}. Despite its merits, it can be computationally very expensive, especially in context of robot planning and control, since (a) unlike in process industries, typical robotic systems demand re-planning online at high frequency, (b) most systems have a non-linear dynamical model and (c) constraints apply both on state and controls. Hence, the nonlinear-MPC (NMPC) poses a number of challenges in practical implementation \cite{c15}. Lighter variants of MPC such as LMPC, explicit MPC \cite{c15} and other simplified NMPC-based methods \cite{c15} have emerged. However, LMPC gradually induces uncertainties and fails for highly non-linear systems where the range of linearization is narrow and inadequate \cite{c6}. Explicit MPC is not practical for higher state and input states due to expensive memory requirements \cite{c15}. %Other MPC based methods either address only a part of the problem or are intended for specific applications.  
In \cite{c16}, the authors proposed a decoupling principle under a small noise assumption and demonstrated first order  near optimality of the decoupled control law for general non-linear systems.\\ %The same principle is also extended to partially observed systems \cite{c17}. 

This paper establishes a decoupling principle that consists of a nominal open loop controls sequence, along with a precisely defined linear feedback law dependent on the open loop, derived using a perturbation expansion of the Dynamic Programming (DP) equation, that is near optimal to the third order, and hence, can work for even moderate noise levels. Further, we perform an extensive empirical comparison of our proposed technique, the ``Trajectory optimized Perturbation Feedback Control (T-PFC)", with the NMPC technique, that shows near identical performance up to moderate noise levels, while taking approximately as much as $100\times$ less time than NMPC to execute in some examples (cf. Fig. 1 and Table I). \\

\section{PROBLEM FORMULATION AND PRELIMINARIES}
This section outlines the details of the system considered and the problem statement.
\subsection{System description}
Let $x_{t} \in \mathcal{X} \subset{\mathbb{R}^{n_x}}$ and $u_{t} \in \mathcal{U}  \subset{\mathbb{R}^{n_u}}$ denote the system state and control input at time $t$ respectively, with $\mathcal{X}$ and $\mathcal{U}$ being corresponding vector spaces. We consider a control-affine nonlinear state propagation model with $f:\mathcal{X} \to \mathcal{X}$ and $g:\mathcal{X} \to \mathcal{X}$ as, 
$
{\bf x_{t+1}} = f({\bf x_t}) + g({\bf x_t}){\bf u_t + \epsilon \sqrt{dT} \omega_t}$, where, $ {\bf \omega_t} \in  \mathcal{N}(0, {\bf I}) $ is an i.i.d zero mean Gaussian noise with variance ${\bf I}$, $dT$ is the discretization time for the continuous time Stochastic Differential Equation (SDE) $d{\bf x} = \bar{f}(\bf x)dt +\bar{g}(\bf x)u dt + \epsilon dw$, and $\epsilon$ is a scaling factor. 
In particular, the discrete time dynamics are obtained from the SDE as follows:
$f(\bf x_t) = x_t + \bar{f}({\bf x_t})dT, \, g({\bf x_t}) = \bar{g}({\bf x_t})$ and the noise term becomes $\epsilon \sqrt{dT} {\bf \omega_t}$, where $\omega_t$ are standard Normal random variables. The reason we explicitly introduce the discretization time $dT$ will become clear later in this section. It is assumed from hereon that $O(dT^2)$ terms are negligible, i.e, the discretization time is small enough.

\subsection{Stochastic optimal control problem }
Given an initial state ${\bf x_0}$, the problem of stochastic optimal control \cite{c18}, in this case, for a fully observed system, is to solve  
\begin{align}
\mathop{min}_{\pi}\,  \mathop{\mathbb{\Huge E}}_{\omega_t} \Big{\lbrack} C_N({\bf x_N}) + \sum_{t= 0}^{N-1} C_t({\bf x_t, u_t})\Big{\rbrack}\\
s.t \quad  {\bf x_{t+1}} = f({\bf x_t}) + g({\bf x_t}){\bf u_t + \epsilon \sqrt{dT} \omega_t} \nonumber
\end{align}
 for a sequence of admissible control policies $\pi = \{\pi_0, \pi_1,..\pi_t,.,\pi_{N-1} \}$, where $\pi_t:\mathbb{\mathcal{X} \to \mathcal{U}}$, \(C_t:\mathcal{X} \times \mathcal{U} \to \mathcal{R}\) denotes the incremental cost function and \(C_K:\mathcal{X} \to \mathcal{R} \), the terminal cost.
 
\subsection{Definitions}
Let $({\bf \bar{x}_t, \bar{u}_t})$ represent the nominal trajectory  
of the system, with its state propagation described by the model, $ {\bf\bar{x}_{t+1}} = f({\bf \bar{x}_t}) + g({\bf \bar{x}_t}){\bf \bar{u}_t}$. Let $(\delta {\bf x_t}, \delta {\bf u_t})$ denote the perturbation about its nominal, defined by $\delta {\bf x_t} = {\bf x_t} - {\bf \bar{x}_t}, \delta {\bf u_t} = {\bf u_t} - {\bf \bar{u}_t}$. Now, by Taylor's expansion of (1) about the nominal $({\bf \bar{x}_t, \bar{u}_t})$ and the zero mean ${\bf w_t}$, the state perturbation can be written as \(\delta {\bf x_{t+1}} =  A_t {\bf \delta x_t} + B_t {\bf \delta u_t} + {\bf \epsilon \sqrt{dT} \omega_t} + r_t\), where $A_t= \frac{\partial f({\bf x_t})}{\partial {\bf x_t}}\arrowvert_{\bf \bar{x}_t} + \frac{\partial g({\bf x_t})}{\partial {\bf x_t}}\arrowvert_{\bf \bar{x}_t} {\bf \bar{u}_t}$, $B_t = g({\bf \bar{x}_t})$ and $r_t$ represents higher order terms. 

Let $\bar{J}_t({\bf x_t})$ denote the optimal cost-to-go function at time $t$ from ${\bf x_t}$ for the deterministic problem (i.e, $\epsilon =0$), and $\bar{J}_t^{\epsilon}({\bf x_t})$ denote the optimal cos-to-go function of the stochastic problem.  We expand the deterministic cost-to-go quadratically about the nominal state in terms of state perturbations as $\bar{J}_t({\bf x_t}) = \bar{J}_t({\bf \bar{x}_t}) + G_t \delta {\bf x_t} + \frac{1}{2} \delta {\bf x_t^\intercal} P_t \delta {\bf x_t}+ q_t$, where, $G_t = \frac{\partial \bar{J}_t({\bar x_t})}{\partial {\bf x_t}}^\intercal\arrowvert_{\bf \bar{x}_t} $, $P_t = \frac{\partial^2 \bar{J}_t({\bar x_t})}{\partial^2 {\bf x_t}}\arrowvert_{\bf \bar{x}_t}$ and $q_t$ denotes the higher order terms. 

Finally, we consider a step cost function of the form 
$C_t({\bf x_t, u_t}) = l({\bf x_t}) + \frac{1}{2} {\bf u_t^\intercal} R {\bf u_t}$ and let $L_t = \frac{\partial l({\bf x_t})}{\partial {\bf x_t}}\arrowvert_{\bf \bar{x}_t}$ and $L_{tt} = \frac{\partial^2 l({\bf x_t})}{\partial^2 {\bf x_t}}\arrowvert_{\bf \bar{x}_t} $.
%\subsection{Assumptions}
Using the definitions above, we assume that the functions $f({\bf x_t})$, $\bar{J}({\bf x_t})$ and $l({\bf x_t})$ are sufficiently smooth over their domains such that the requisite derivatives exist and are uniformly bounded.

\section{A NEAR OPTIMAL DECOUPLING PRINCIPLE}
\begin{figure}[thpb]
      \centering
    \subfloat[]{{\includegraphics[width=0.7\linewidth]{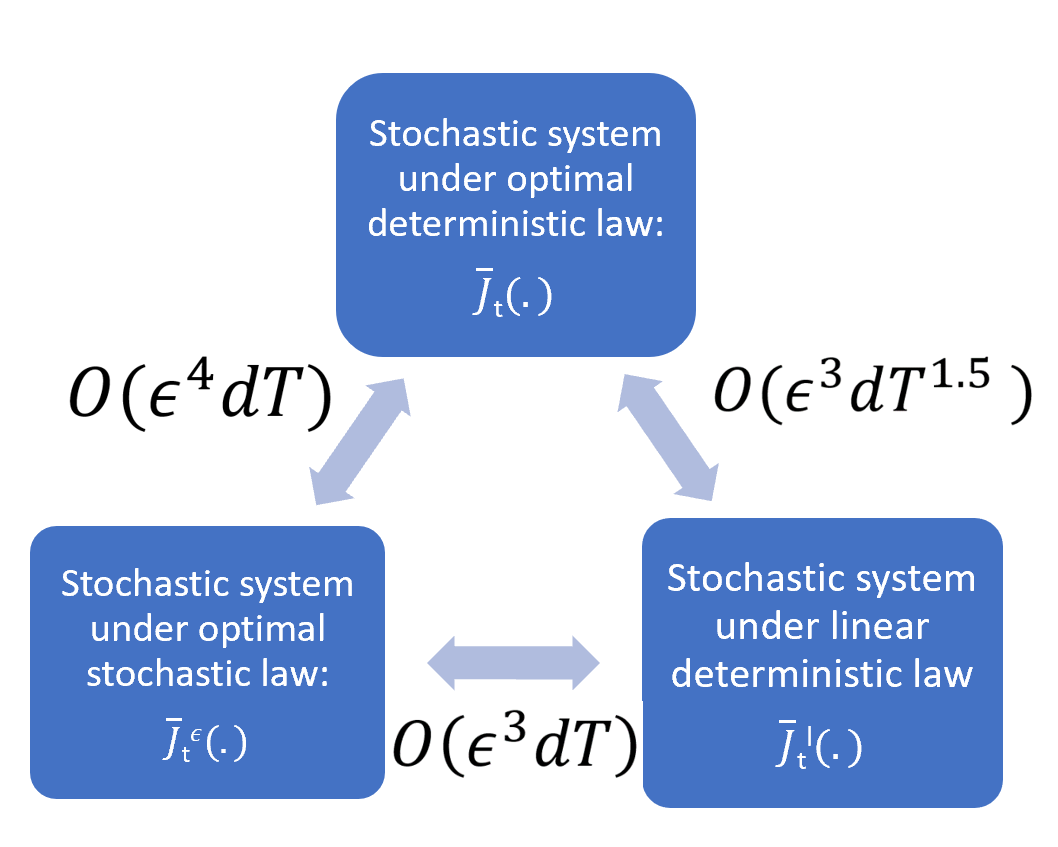}}}
    \caption{\small Schematic of the Near-Optimal Decoupling Principle}%
    \label{NOD}
   \end{figure}

This section states a near-optimal decoupling principle that forms the basis of the T-PFC algorithm presented in the next section. Our program in this section shall be as follows:
\begin{itemize}
    \item \textit{Decoupling:} First, we shall show that the optimal open loop control sequence of the deterministic problem (given by the gains $G_t$) can be designed independent of the closed loop gains determined by $P_t$, i.e, the $P_t$ do not affect the $G_t$ equations for an optimal control sequence in the deterministic problem.
    \item \textit{Step A:} Next, we shall only keep the first two terms in the optimal deterministic feedback law, i.e., ${\bf u_t^{l}}= \bar{\bf u_t} + K_t \delta {\bf x_t}$, and show that the closed loop performance of the truncated linear law is within $O(\epsilon^3 dT^{3/2})$ of the full deterministic feedback law when applied to the \textit{stochastic system}.
    \item \textit{Step B:} Finally, we will appeal to a result by Fleming \cite{fleming} that shows that the closed loop performance of the full deterministic law applied to the stochastic system is within $O(\epsilon^4 dT)$ of the optimal stochastic closed loop, and show that the stochastic closed loop performance  of the truncated linear feedback law is within $O(\epsilon^3dT)$ of the optimal stochastic closed loop
\end{itemize}
The scheme above is encapsulated in Fig. \ref{NOD}.

{\bf Proposition 1:} \textbf{Decoupling.}
Given an optimal nominal trajectory, the backward evolutions of the deterministic gain $G_t$ and the covariance $P_t$ of the optimal cost-to-go function $\bar{J}_t({\bf x_t})$, initiated with $G_N = \frac{\partial \bar{C}_N({\bar x_N})}{\partial {\bf x_N}}^\intercal\arrowvert_{\bf \bar{x}_N} $ and $P_N = \frac{\partial^2 \bar{C}_N({\bar x_N})}{\partial^2 {\bf x_N}}\arrowvert_{\bf \bar{x}_N}$ respectively, are as follows: \\
\begin{equation}
\hspace*{-5.72cm}G_t = L_t + G_{t+1}A_t \label{OL}
\end{equation}
\begin{equation}
\hspace*{-1.5cm}P_t = L_{tt} + A_t^\intercal P_{t+1} A_t - K^\intercal_t S_t K_t + G_{t+1} \otimes \tilde R_{t,xx}  \label{feedback}
\end{equation}
for $t = \{0,1,...,N-1\}$, where, $S_t = (R_t + B_t^\intercal P_{t+1}B_t), K_t = -S_t^{-1}(B_t^\intercal P_{t+1} A_t + (G_{t+1} \otimes \tilde{R}_{t_{x u}})^\intercal), \tilde{ R}_{t,xx} = \nabla^{2}_{xx}f({\bf x_t})\arrowvert_{{\bf \bar{x}_t}} + \nabla^{2}_{xx}g({\bf x_t})\arrowvert_{{\bf \bar{x}_t}} {\bf \bar{u}_t}, \tilde{ R}_{t,xu} = \nabla^{2}_{xu}(f({\bf x_t}) + g({\bf x_t}){\bf {u}_t}) \arrowvert_{\bf \bar{x}_t, \bar{u}_t}$ where $\nabla^2_{xx}$ represents the Hessian of a vector-valued function w.r.t $x$ and $\otimes$ denotes the tensor product. 

Proof for the above is provided in the appendix section. In essence, the key step in the proof of proposition-1 is in realizing that when the nominal trajectory is optimal, the term corresponding to the open-loop control trajectory vanishes in deriving an expression for perturbed control as shown in equation (4) and thereafter. This means that the dependency of the perturbed variables in the design of the nominal trajectory is nullified resulting in equations (2) and (3). It may be noted here that equation (2) corresponds to the co-state equation following the first order optimality conditions over an optimal nominal trajectory, whereas equation (3) is a discrete time dynamic Riccati-like equation dictating the feedback law design. The consequence of the above result is that the second order sensitivity matrix in the expansion of the cost, $P_t$ which determines the feedback gain $K_t$, doesn't influence the first order sensitivity matrix $G_t$ (the co-state) that determines the optimal open-loop sequence. Thus, the decoupling between the nominal and linear feedback holds true. In other words, the design of an optimal control policy in a fully-observed problem as in (1) can be decoupled into the design of an open-loop deterministic nominal (${\bf \bar{x}_t}$, ${\bf \bar{u}_t}$) and then a linear feedback law whose coefficients can be extracted through a time-recursive propagation of (2) and (3), but which is nonetheless near optimal to third order ($O(\epsilon^3 dT)$) as we shall show below.\\
\textbf{Step A.} Let the optimal deterministic feedback law  for the deterministic system ($\epsilon=0$) be given by: ${\bf u_t}({\bf x_t}) = \bar{\bf u_t} + K_t \delta {\bf x_t} + R(\delta {\bf x_t})$. The result above gives us the recursive equations required to solve for ${\bf \bar{u}_t}$ in terms of $G_t$, and $K_t$ in terms of $P_t$. Consider the truncated linear feedback law, i.e., ${\bf u_t^{l}}({\bf x_t}) = {\bf \bar{u}_t} + K_t\delta {\bf x_t}$. Now, we shall apply the control laws ${\bf u_t(.)}$ and ${\bf u_t^{l}(.)}$ to the stochastic system ($\epsilon \neq 0$) and compare the closed loop performance. It can be shown that the state perturbations from the nominal under the optimal deterministic law evolve according to $\delta {\bf x_{t+1}} = \bar{A}_t \delta {\bf x_t} + B_t R(\delta {\bf x_t}) + S_t(\delta {\bf x_t}) + \epsilon \sqrt{dT} \omega_t$,  while that under the truncated linear law evolves according to $\delta {\bf x_{t+1}^{l}} = \bar{A}_t \delta {\bf x_t^{l}} + S_t(\delta {\bf x_t^l})+\epsilon \sqrt{dT} \omega_t$, where $\bar{A}_t = A_t + B_t K_t$ is the linear closed loop part, and $S_t(.)$ are the second and higher order terms in the dynamics. The closed loop cost-to-go under the full deterministic feedback law is then given by: $\bar{J}_k({\bf x_k}) = E[\sum_{t=k}^N c({\bf \bar{x}_t}, {\bf \bar{u}_t}) + C^1_t \delta {\bf x_t} + \delta {\bf x_t^{'}} C^2_t\delta {\bf x_t} + H_t(\delta {\bf x_t})]$, and that for the truncated linear law is given by: $\bar{J}_k^{l}({\bf x_k}) = E[\sum_{t=k}^N c({\bf \bar{x}_t}, {\bf \bar{u}_t}) + C^1_t \delta {\bf x_t^{l}} + \delta {\bf x_t^{l}}^\intercal C^2_t\delta {\bf x_t^{l}} + H_t(\delta {\bf x_t^{l}})]$, where $C_t^1$, $C_t^2$, are the first and second order coefficients of the step cost expansion that depend only on the nominal $({\bf \bar{x}_t}, {\bf \bar{u}_t)}$, and $H_t(.)$ denote third and higher order terms of the expansions. Then $\bar{J}_k({\bf x_k}) - \bar{J}_k^{l}({\bf x_k}) =\sum_{t=k}^N \underbrace{E[\delta {\bf x_t^{'}}C_t^2\delta {\bf x_t} - \delta {\bf x_t^{l^\intercal}}C^2_t \delta {\bf x_t^{l}}]}_{T_1} +\\
\underbrace{\sum_{t=k}^N E[H_t(\delta {\bf x_t}) - H(\delta {\bf x_t^{l}})]}_{T_2}$, by noting that $\sum_{t=k}^N C^1_t\delta {\bf {x}_t}=0$ and $\sum_{t=k}^N C^1_t \delta {\bf {x}_t^{l}} = 0$ (from the first order condition for a minimum at the optimal nominal trajectory).
%, by noting that $E[\delta x_t]=0, E[\delta x_t^{l}]=0$.
Consider the deviation between the two closed loops $\delta {\bf x_t} - \delta {\bf x_t^{l}} = \bar{A}_t(\delta {\bf x_t} - \delta {\bf x_t^{l}}) + B_t R_t(\delta {\bf x_t}) + S_t(\delta {\bf x_t}) -S_t(\delta {\bf x_t^l})$, where note that $||R_t(\delta {\bf x_t})|| = O(\epsilon^2 dT)$, as are $||S_t(\delta {\bf x_t^l})||$ and  $||S_t(\delta {\bf x_t})||$ since they consist of second and higher order terms in the feedback law and the dynamics respectively, when $\epsilon \sqrt{dT}$ is small. Therefore, it follows that the closed loop state deviation between the full deterministic and the truncated linear law is $||\delta {\bf {x_t}} - \delta {\bf x_t^{l}}|| = O(\epsilon^2dT)$. Further, it is also true that $\delta {\bf x_t}$ and $\delta {\bf x_t^{l}}$ are both $O(\epsilon \sqrt{dT})$. Hence, using the above it follows that both terms $T_1$ and $T_2$ are $O(\epsilon^3dT^{3/2})$. \textit{Therefore, it follows that the difference in the closed loop performance of the full deterministic feedback law and the truncated linear feedback law is $|\bar{J}_k({\bf x_k}) - \bar{J}_k^{l}({\bf x_k})|= O(\epsilon^3 dT^{3/2})$.}\\
From the above discussion, it is evident that the third order optimality is primarily resulted from the linear feedback law that adapts to the nominal cost and is exact. Hence, it follows that with other approximate or inexact linear feedback laws, as in for instances - T-LQR and ILQG/DDP, the performance remains $O(\epsilon^2 dT)$ optimal.\\
\textbf{Step B:} Now, we shall establish the closeness of the optimal stochastic closed loop and the stochastic closed loop under the truncated linear feedback law.
First, we recount a seminal result due to Fleming \cite{fleming} regarding the "goodness" of the deterministic feedback law for the stochastic system. Fleming considered the continuous time SDE: $d{\bf x} = \bar{f}({\bf x}) dt + g({\bf x}){\bf u}dt + \epsilon d{\bf w}$. Let the cost-to-go of the optimal stochastic closed loop be given by $\bar{J}^{\epsilon}(t,{\bf x})$, and let the cost-to-go of the closed loop under the deterministic law be given by $\bar{J}(t,{\bf x})$. Then, it is shown that the functions $J^{\epsilon}$ and $J$ have the following perturbation expansion in terms of $\epsilon$: $J^{\epsilon} = \varphi + \epsilon^2 \theta + \epsilon^4 \chi$,and $J^0 = \varphi + \epsilon^2 \theta + \epsilon^4 \chi'$, where $\varphi$, $\theta$,$\chi$ and $\chi'$ are functions of $(t,{\bf x})$. Therefore, it follows that the difference in the closed loop performance between the optimal stochastic and optimal deterministic law \textit{when applied to the stochastic system} is $O(\epsilon^4)$!\\
If we adapt this result to our discrete time case with a time discretization $dT$, where $O(dT^2)$ is negligible, then the difference between the true stochastic closed loop performance and that under the deterministic optimal law, $|J^{\epsilon}_t({\bf x_t}) - J_t({\bf x_t})| = O(\epsilon^4dT)$. Thus, using the above result and the result form step A, it follows that \textit{difference between the closed loop performance of the truncated linear feedback law and that of the otpimal stochastic closed loop, $|J_t^{\epsilon}({\bf x_t}) - J_t^{l}({\bf x_T})| = O(\epsilon^3dT)$ at the least.} This establishes the near optimality of the truncated linear feedback closed loop.\\
\textbf{ILQG/DDP:} The condition in (\ref{OL}) is precisely when the ILQG/ DDP algorithms are deemed to have converged. However, that does not imply that the feedback gain at that stage for ILQG/ DDP is the same as that in Eq. (\ref{feedback}), and in fact, the feedback gains of ILQG/ DDP are different from that in Eq. \ref{feedback} as we shall see in our examples. The basic idea in the development above is to design an open loop optimal sequence, and then design a feedback gain according to Eq. \ref{feedback}, it is in this second step that we differ from ILQG/ DDP (which are methods to get open loop optimal sequences and make no claims about the feedback gains).

\section{Trajectory-Optimized Perturbation Feedback Control (T-PFC)}
In this section, we formalize the Trajectory-optimized Perturbation Feedback Control (T-PFC) method based on the decoupling principle of the previous section. 

\subsection{Nominal Trajectory Design}
The optimal nominal trajectory can be designed by solving the deterministic equivalent of problem (1), which can be formulated as an open-loop optimization problem as follows:
\begin{align*}
\mathop{min}_{{\bf \check{u}}}\,  \Big{\lbrack} C_N({\bf x_N}) + \sum_{t= 0}^{N-1} C_t({\bf x_t, u_t}) \Big{\rbrack}\\
s.t \quad  {\bf x_{t+1}} = f({\bf x_t}) + g({\bf x_t})\bf u_t
\end{align*}
where, ${\bf \check{u}} =\{{\bf u_0}, {\bf u_1},..{\bf u_{N-1}}\}$. This is a design problem that can be solved by a standard NLP solver. The resultant open-loop control sequence together with a sequence of states obtained through a forward simulation of noise-free dynamics constitute the nominal trajectory.\\
Constraints on the state and the control can be incorporated in the above problem as follows:\\ 
{\bf State constraints:}
Non-convex state constraints such as obstacle avoidance can be dealt by imposing exponential penalty cost as barrier functions. Obstacles can be circumscribed by Minimum Volume Enclosing Ellipsoids (MVEE)\cite{c17} that enclose a polygon given its vertices. Such kind of barrier functions can be formulated by \cite{c19}: 
$C_{obs}({\bf x_t}) = \sum_{m=1}^{n} \Gamma_m \exp(-\rho_m({\bf x_t} - c^{m})^\intercal \mathcal{E}^{m} ({\bf x_t} - c^{m}) )$,
where, $c^{m}$ and $\mathcal{E}$ correspond to the center and geometric shape parameters of the $m^{th}$ ellipsoid respectively. $\Gamma_m$ and $\rho_m$ are the scaling factors. Obstacles are assimilated into the problem by adding $C_{obs}({\bf x_t})$ to the incremental cost $C_t({\bf x_t}, {\bf u_t})$.\\
{\bf{Control bounds:}} Control bounds can safely be incorporated while designing the optimal nominal trajectory as hard constraints in the NLP solver. In this case, the constraints are linear in control inputs and the modified incremental cost function can be written as $C_t'({\bf x_t},{\bf u_t}) = C_t({\bf x_t},{\bf u_t}) + \mu_t ({F_t{\bf u_t} + H_t})$. The first order condition (4) is then modified to 
$R_t {\bf \bar{u}_t} + B_{t}^\intercal G_{t+1}^\intercal + F_t^\intercal\mu_t^\intercal = 0$ using KKT conditions \cite{c20}, which upon utilizing in the derivation of expression for $\delta {\bf u_t}$ nullifies the influence of $\mu_t$. Hence, equations (3), (4) and (6) will remain the same. 

\subsection{Linear Feedback Controller Design}
Given a nominal trajectory $({\bf \bar{x}}, {\bf \bar{u}})$, a linear perturbation feedback controller around it is designed by pre-computing the feedback gains. The sequence of $K_t$ is determined by a backward pass of $G_t$ and $P_t$ as described by (3) and (4). The linear feedback control input is given by $\delta {\bf u_t}= K_t \delta {\bf x_t}$. Hence, $ {\bf u_t} = {\bf \bar{u}_t} + \delta {\bf u_t}= {\bf \bar{u}_t} + K_t ({\bf x_t} - {\bf \bar{x}_t})$
forms the near-optimal online control policy. Algorithm-1 outlines the complete T-PFC algorithm.\\
{\bf Complexity}: The computational complexity of the deterministic open-loop optimal control problem is O($b N n_x^2$), assuming $b$ iterations in obtaining a valid solution. Solving the recursive equations concerning the feedback is O($N n_x^3$). Hence, the total complexity of T-PFC is O($b N n_x^2 + N n_x^3$).
\begin{algorithm}
  \caption{\strut T-PFC}
  {\bf Input:} Initial State - ${\bf x_0}$, Goal State - {$\bf{x_f}$}, Time-step $\Delta t$, Horizon - $N$, System and environment parameters - $\mathcal{P}$\;	
  $t \gets 0$\;
  \While {$\|{\bf x_t} - {\bf x_f}\| < \varepsilon$}{
  	\If {Cost fraction$ > c_{th}$ \textnormal{or} t == 0 \textnormal{or} t == N-1}{
   {$({\bf \bar{x}_{t:N-1}}$, ${\bf \bar{u}_{t:N-1}})\leftarrow$Plan(${\bf x_t}, \mathcal{P}, {\bf x_f}$)}\\
  Compute parameters:{$\{P_{t:N-1},G_{t:N-1},K_{t:N-1}\}$}  
  }	
  {Policy evaluation: ${\bf u_t} \leftarrow {{\bf \bar{u}_t} + K_t ({\bf x_t} - {\bf \bar{x}_t})}$} \\
  {Process: ${\bf x_{t+1}} \leftarrow f({\bf x_t}) + g({\bf x_t}){\bf u_t + \epsilon \omega_t}$}
  $t \leftarrow t + 1$ 
  }
\end{algorithm}

\section{EXAMPLE APPLICATIONS AND RESULTS}

This section demonstrates T-PFC in simulation with three examples. The $Gazebo$ \cite{c21} physics engine is used as a simulation platform in interface with $ROS$ middleware\cite{c22}. Numerical optimization is performed using the $Casadi$ \cite{c23} framework employing the $Ipopt$ \cite{c24} NLP software. A feasible trajectory generated by the non-holonomic version of the RRT algorithm \cite{c25} is fed into the optimizer for an initial guess. Simulations are carried out in a computer equipped with an Intel Core i7 2.80GHz $\times$ 8 processor. The results presented in each example are averaged from a set of 100 Monte Carlo simulations for a range of tolerable noise levels $\epsilon$. The proposed approach has been implemented to the problem of motion planning under process noise in the dynamical model to obtain the cost plots and then simulated in a physics engine on a realistic robot model for further analysis.\\
{\bf Noise characterization:} Process noise is modeled as a standard Brownian noise added to the system model with a standard deviation of $\epsilon \sqrt{dt}$. Since it is assumed to be additive Gaussian and i.i.d (even w.r.t the noise in other state variables), it could account for various kinds of uncertainties including that of parametric, model and the actuator. $\epsilon$ is a scaling parameter that is varied to analyze the influence of the magnitude of the noise. Other case-specific parameters are provided in Table II.

For simulating in physics engine, we use realistic physical robot models in an obstacle-filled environment along with moderate contact friction ($\mu = 0.9$) and drag, which are unaccounted for in our design. Apart from this model uncertainty, we also introduce actuator noise through an additive Gaussian of standard deviation $\epsilon \sigma_t$, where $\sigma_t$ is $\lVert {\bf u_s}\rVert_{\infty}$.

\subsection{Car-like robot}
A 4-D model of a car-like robot with its state described by $(x_t,y_t,\theta_t, \phi_t)^\intercal$ is considered. For a control input constituting of the driving and the steering angular velocities, $(u_t, w_t)^\intercal$, the state propagation model is as follows:
\begin{align*}
\dot{x} &= u cos(\theta), \hspace*{.5cm} \dot{\theta}= \frac{u}{L} tan(\phi)\\
\dot{y} &= u sin(\theta), \hspace*{.5cm}  \dot{\phi}= \omega \\
\end{align*}
Fig. 4 shows the path taken by a car-like robot in an environment filled with 8 obstacles enclosed in MVEEs. Plots in Fig. 3 (a) indicate the averaged magnitude of both the nominal and the total control signals at $\epsilon = 0.25$. The standard deviation of the averaged total control sequence, in both plots, from the nominal is less than one percent of it.  
\begin{figure}[thpb]
      \centering
      \subfloat[]{\includegraphics[width=0.5\linewidth]{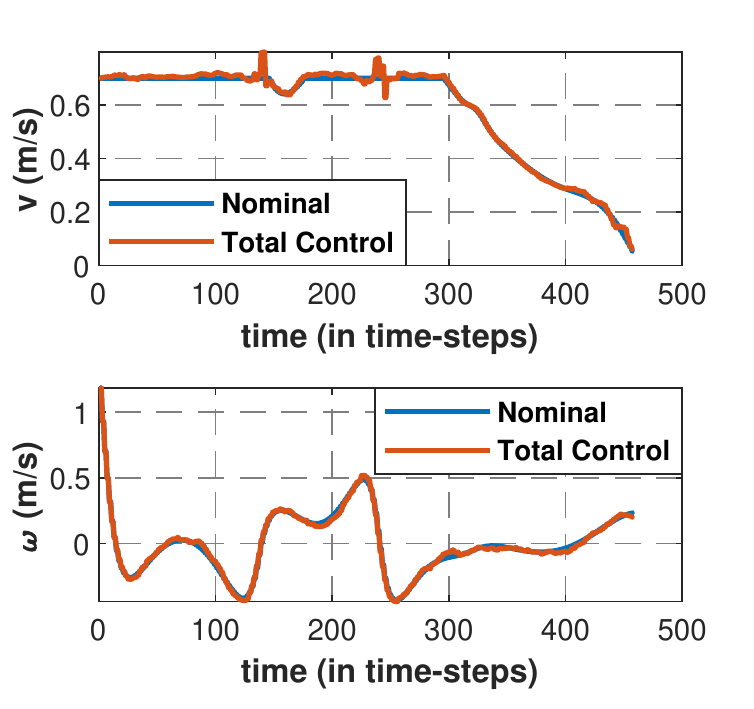}}
      \subfloat[]{\includegraphics[width=0.5\linewidth]{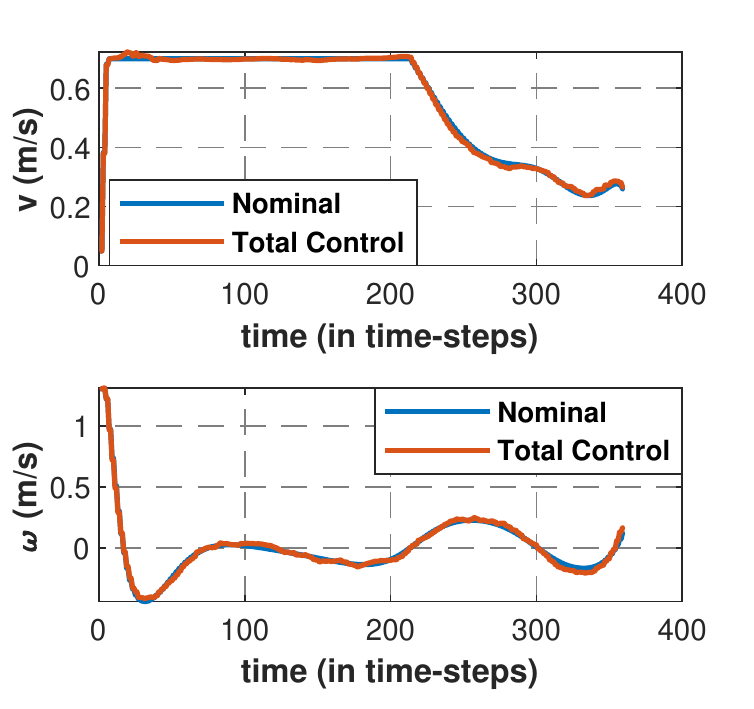}}
      \caption{\small{Optimal nominal and total control inputs (averaged) at $\epsilon$ = 0.25 for (a) a car-like robot and (b) car with trailers}}
      \label{controls_car}
\end{figure}
\begin{figure}[thpb]
      \centering
      \subfloat[Rviz trajectory]{{\includegraphics[width=0.3\linewidth]{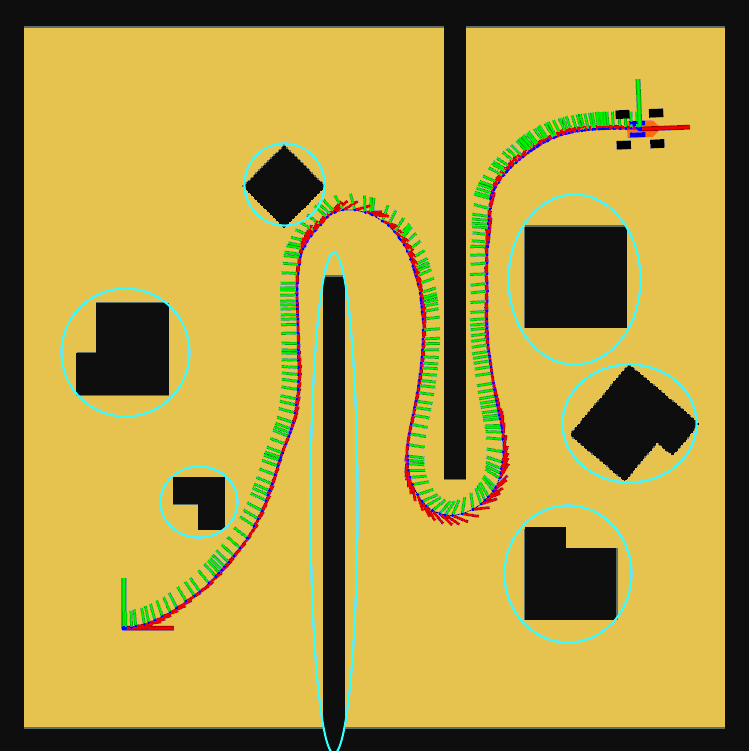} }}%
    %\qquad
    \subfloat[Robot's world in Gazebo]{{\includegraphics[width=0.37\linewidth]{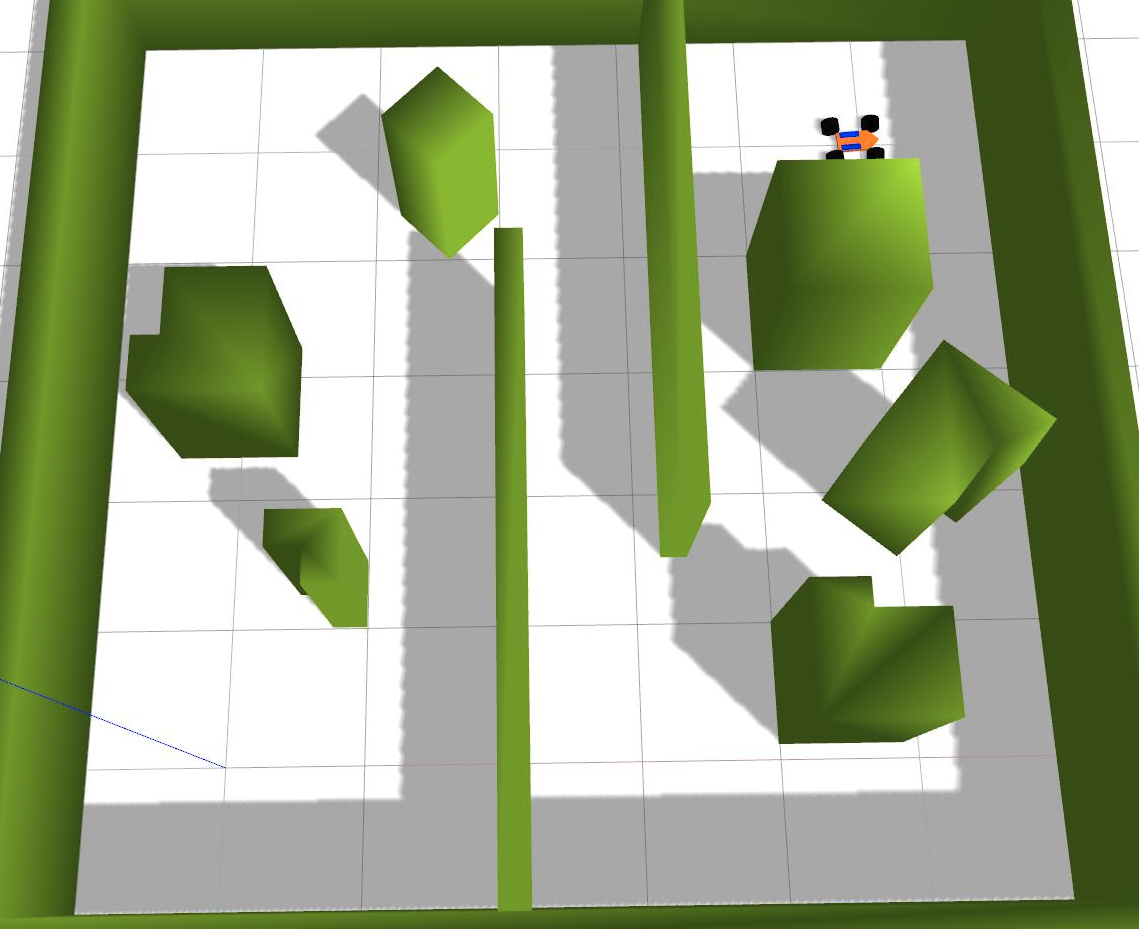} }}%
    \caption{\small{Motion Planning of a car-like robot using T-PFC for an additive control noise of standard deviation = 25\% of the norm of saturation controls $i.e,$  $\epsilon=0.25$}}%
    \label{example_1}%
   \end{figure}
\subsection{Car-like robot with trailers}
With $n$ trailers attached to a car-like robot, the state of a car-like-robot is augmented by $n$ dimensions, each additional entry describing the heading angle of the corresponding trailer. In the current example, $n$ = 2 trailers are considered and their heading angles are given by \cite{c26}:
\begin{align*}
\dot{\theta}_1 &= \frac{u}{L} sin(\theta - \theta_1)\\
\dot{\theta}_2 &= \frac{u}{L} cos(\theta - \theta_1)sin(\theta_1 - \theta_2)
\end{align*}
Hence, the robot has six degrees of freedom. Its world is considered to be composed of four obstacles as shown in Fig. 5. The robot, its environment and its trajectory shown are at $\epsilon=0.25$. Fig. 5(b) displays the nominal and the total control signals averaged at $\epsilon = 0.25$.
   \begin{figure}[thpb]
      \centering
      \subfloat[Rviz trajectory]{{\includegraphics[width=0.3\linewidth]{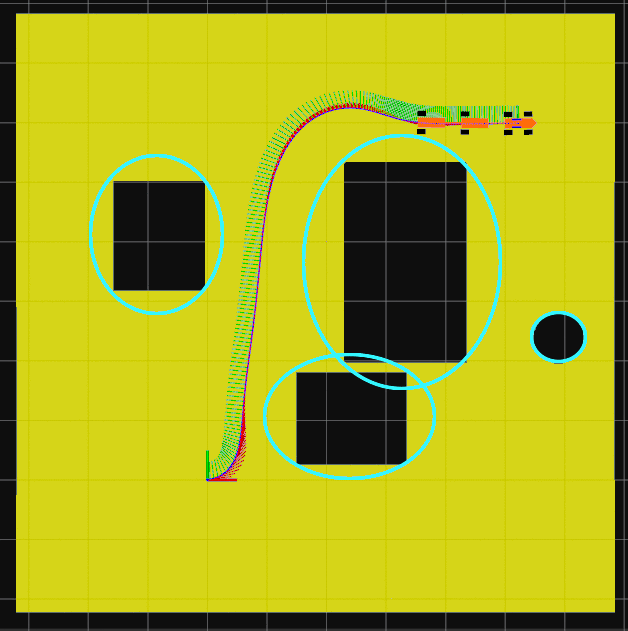} }}%
    %\qquad
    \subfloat[Robot's world in Gazebo]{{\includegraphics[width=0.5\linewidth]{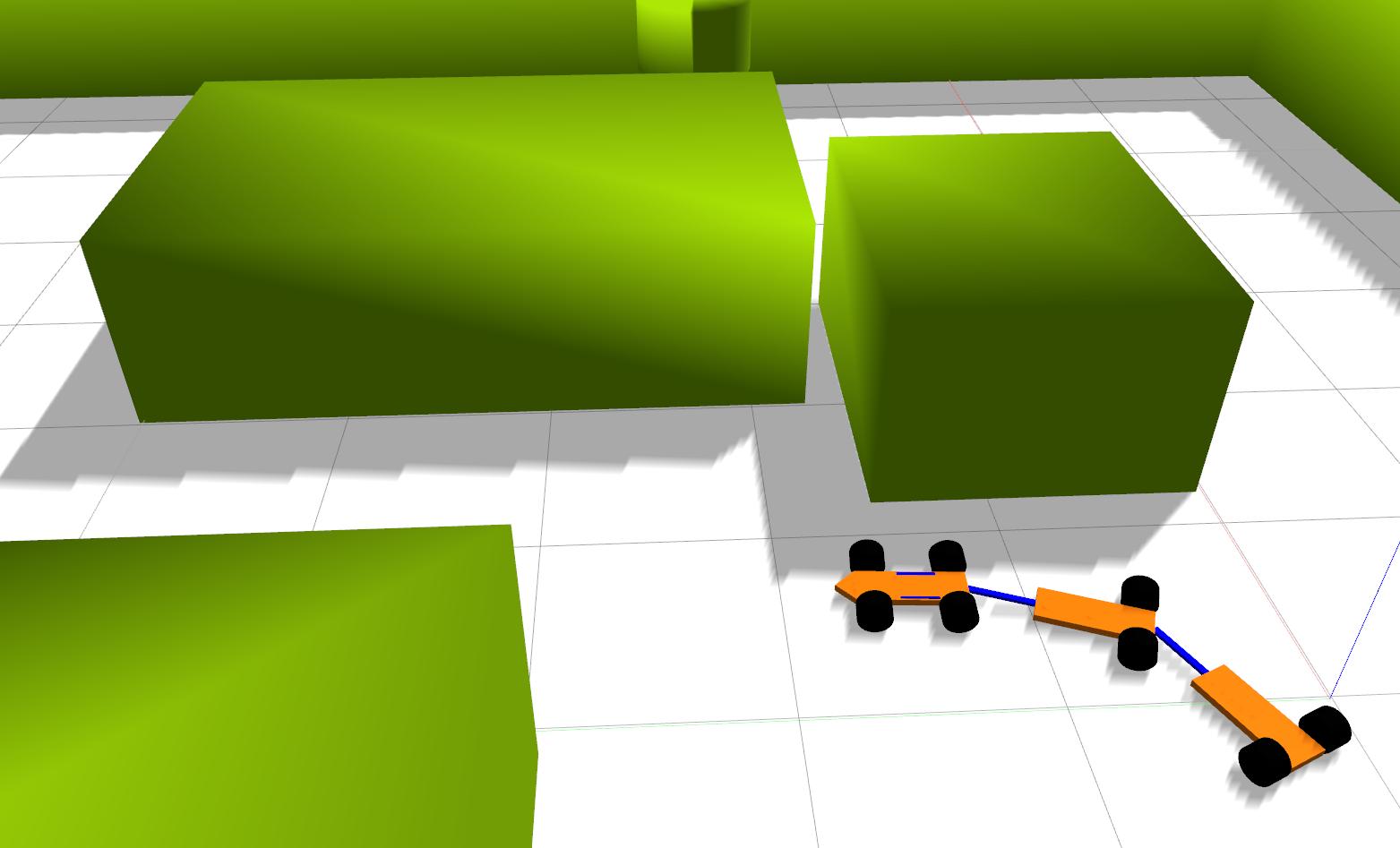} }}%
    \caption{\small{Motion Planning of a car with trailers using T-PFC for an additive control noise of standard deviation = 25\% of the norm of saturation controls $i.e,$  $\epsilon=0.25$}}.%
    \label{example_2}%
   \end{figure}
\subsection{3D Quadrotor}
The 12 DOF state of a Quadrotor comprises of its position, orientation and corresponding rates - $({\bf x_t}, {\bf \theta_t}, {\bf v_t}, {\bf \omega_t})^\intercal$. Forces and torques in its body frame are external inputs in the equations below. However, in the real world (and also in gazebo simulation shown here) the control input is typically fed into the motors. Hence, we consider rotor velocities as the control input, which can be obtained by a linear transformation of forces and torques in body frame.
The state propagation model is then given by the following equations \cite{c30}:
\begin{align*}
{\bf \dot{x}_{t}} &= {\bf v_t}, \hspace*{1cm} {\bf \dot{v}_t} = {\bf g} + \frac{1}{m} (R_{\theta_t} {\bf F_b} - k_d{\bf v_t}) \\
{\bf \dot{\theta_t}} &= J_w^{-1} {\bf \omega_t}, \hspace*{.5cm} {\bf \dot{\omega}_t} = I_c^{-1}{\bf \tau_t}
\end{align*}
Simulations are performed using an AR.drone model\cite{c31} and an environment comprising of a cylindrical obstacle as shown in Fig. 6.
 \begin{figure}[thpb]
      \centering
      \subfloat[]{\includegraphics[width=.7\linewidth]{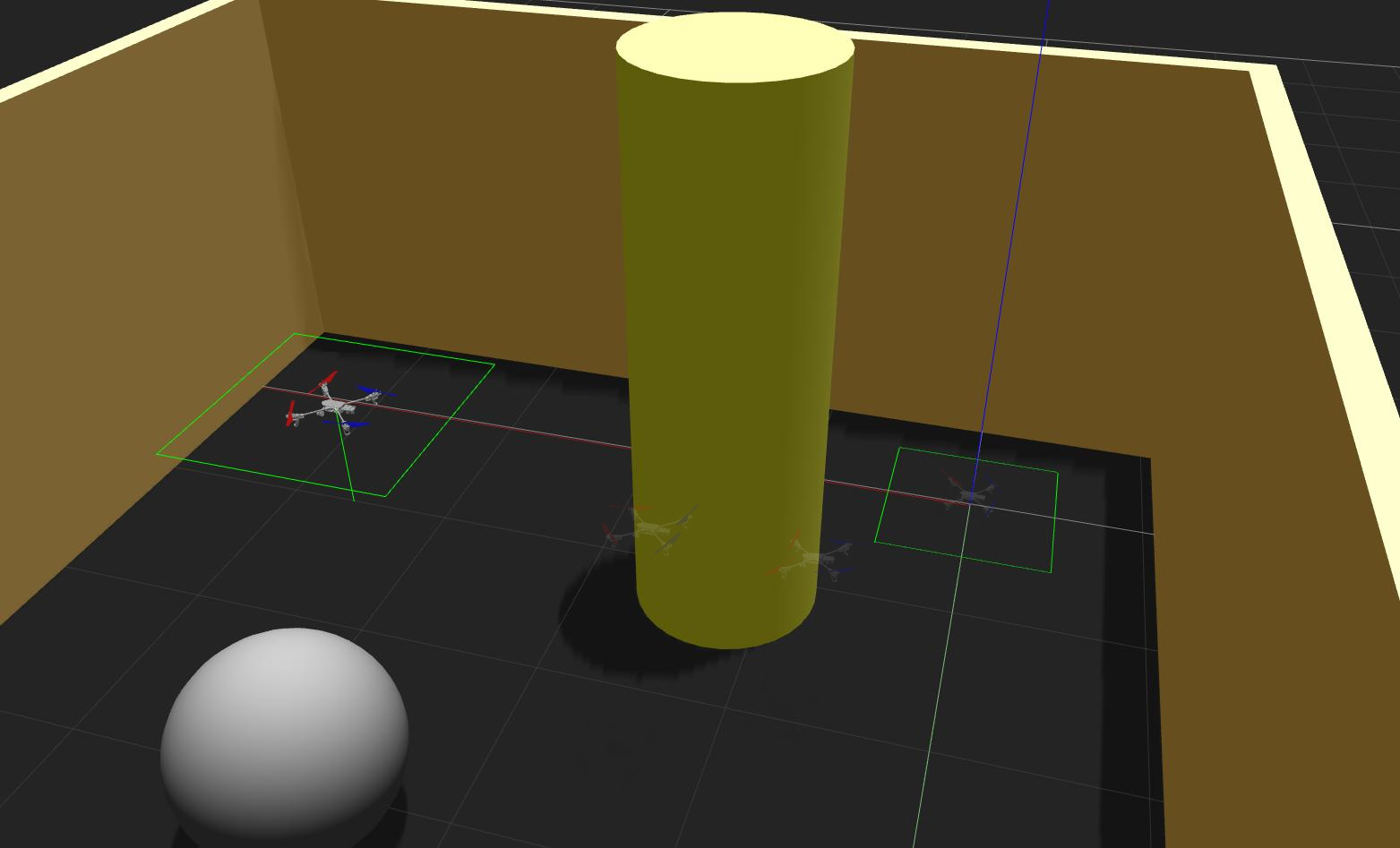}}
      \subfloat[]{\includegraphics[width=.25\linewidth]{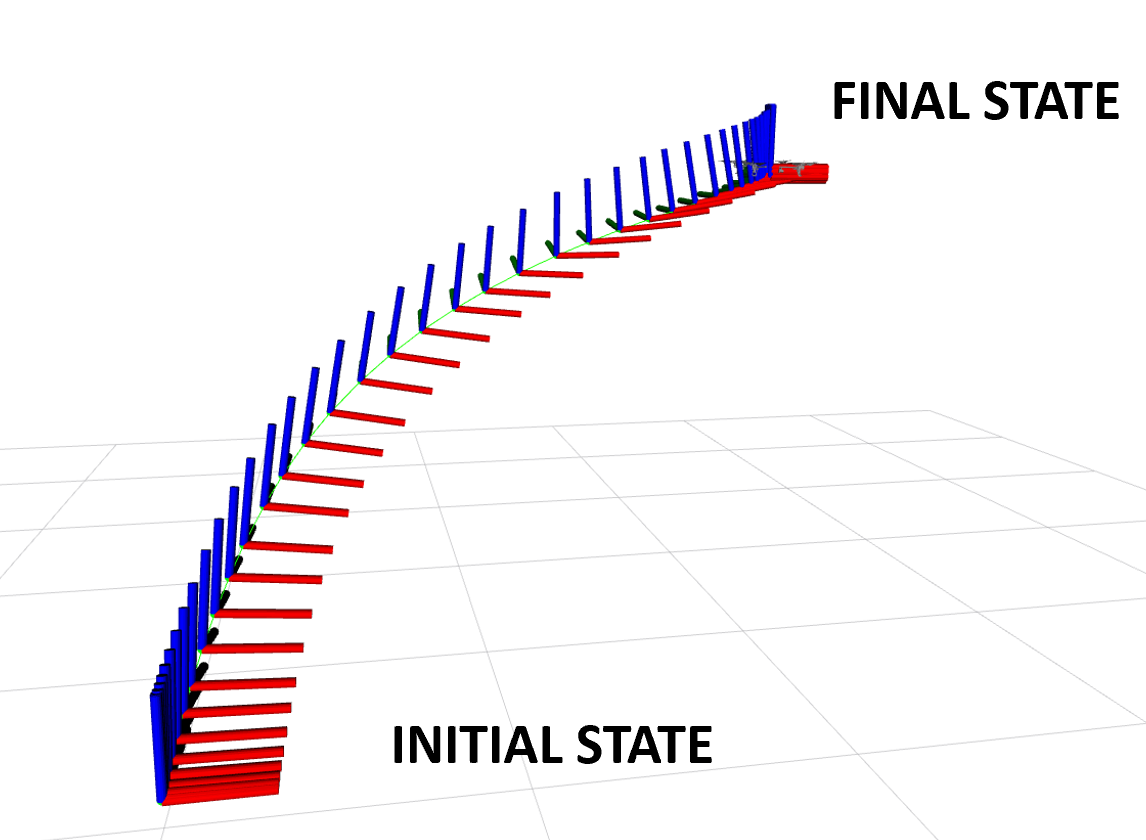}}
      \caption{\small{(a) Quadrotor's world in Gazebo - green boxes represent its initial and final positions respectively (b) trajectory in rviz}}
      \label{quadrotor}
\end{figure}
\begin{table}[h]
\caption{Average run-time of algorithms in seconds}
\label{runtime}
\begin{center}
\begin{tabular}{|c|c|c|c|c|}
\hline
Robot type & MPC & T-LQR & ILQG & T-PFC\\
\hline
Car-like & 447.89 & 4.48 & 161 & 4.52\\
\hline
Car with trailers & 384.42 & 4.11 & 146 & 4.24 \\
\hline
Quadrotor & 71 & 3.33 & 49 & 3.5\\
\hline
\end{tabular}
\end{center}
\end{table}
 \begin{table}[h]
\caption{Simulation parameters}
\label{simulation_parameters}
\begin{center}
\begin{tabular}{|c|c|c|c|}
\hline
 & Car-like & Car with trailers & Quadrotor \\
\hline
 ${\bf x_0}$&$(0,0,0,0)^\intercal$&$(0,0,0,0,0,0)^\intercal$&$(0,0,0.08,0,$\\
 &&&$0,0,0,0,0,0,0,0)'$\\
\hline
 ${\bf x_f}$&$(5,5,0,0)^\intercal$&$(5,6,0,0,0,0)^\intercal$&$(2.6,2.4,1.75,$\\
 &&&$0,0,0)^\intercal$\\
\hline
${N, \Delta t}$& 229, 0.1s & 180, 0.1s & 60, 0.1s\\
\hline
Control& ${\bf u_s^1 =}(0.7,-0.7)$ & ${\bf u_s^1 =}(0.7,-0.7)$& ${\bf u_s^1} = (20,-20)$\\bounds&${\bf u_s^2 =}(-1.3,1.3)$ &${\bf u_s^2 =}(-1.3,1.3)$& ${\bf u_s^i} = (1,-1)
$\\
&&&$ i=2,3,4$ \\
\hline
\end{tabular}
\end{center}
\end{table}
 \begin{figure}[thpb]
      \centering
      \subfloat[]{\includegraphics[width=.5\linewidth]{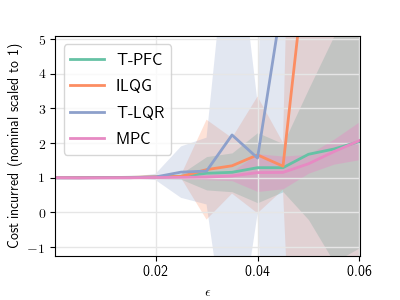}}
      \subfloat[]{\includegraphics[width=.5\linewidth]{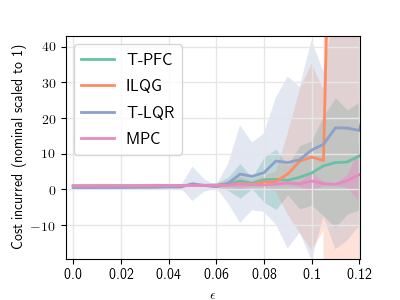}}
      \caption{\small{Cost evolution over a feasible range of $\epsilon$ for (a) car with trailers robot and (b) 3D Quadrotor.}}
      \label{cost_2}
\end{figure}
\subsection{Discussion and comparison of methods}
This section empirically details the implications of the decoupling principle and the T-PFC from the examples in the aforementioned section. Further, we make a comparison here with the Non-linear MPC (NMPC) \cite{c27}, T-LQR \cite{c16} and ILQG \cite{c3}. Average cost incurred, rate of re-planning and time-taken for an execution are chosen as the performance criteria.\\
%, which is same as T-LQR. \\
{\bf Nonlinear MPC:} A deterministic NMPC is implemented with a typical OCP formulation, by re-solving it at every time-step. The NMPC variant implemented here is summarized in  Algorithm-2. The prediction horizon is taken as $N-i$ at the $i^{th}$ time-step. In other words, planning is performed all the way till the end rather than for next few time-steps as in typical MPC. This is done for two reasons: \\
(1) The control sequence obtained this way is equivalent to the deterministic optimal control law that includes higher order terms of feedback control. We wish to juxtapose it with T-PFC that only has a linear feedback (first-order). \\
(2) Due to high penalty cost of multiple barrier functions, the optimizer is prone to failures with smaller prediction horizons. 
Also, by the above arrangement, it follows from Bellman's Principle of Optimality that the predicted open-loop control input will be equal to the optimal feedback policy \cite{c27}. Therefore, this also results in nominal stability.

\begin{algorithm}
  {\bf Input:} Initial State - ${\bf x_0}$, Goal State - {$\bf{x_f}$}, Horizon - $N$, System and environment parameters - $\mathcal{P}$\;	
  $t \gets 0$\;
  \While {$t < N$}{
  	{$({\bf \bar{x}_{t:N-1}}$, ${\bf \bar{u}_{t:N-1}})$ $\leftarrow$ Plan(${\bf x_t},{\bf u_t}, N-t, \bf{x_f},\mathcal{P}$)} \;
  {Process: ${\bf x_{t+1}} \leftarrow f({\bf x_t}) + g({\bf x_t}){\bf \bar{u}_t + \epsilon \omega_t}$}
  $t \leftarrow t + 1$ \;
  }
  \caption{NMPC}
  \end{algorithm}
{\bf T-LQR:} T-LQR is implemented using the same nominal cost as T-PFC. However, the cost parameters of the LQR are tuned entirely separately from the nominal cost \cite{c16}. \\
{\bf ILQG:} ILQG is initiated with the same initial guess as the above three methods. Since the cost contains exponential terms from the barrier functions, it is crucial to carefully choose right parameters for regularization and line search. Regularization is performed by penalizing state deviations in a quadratic modification schedule and an improved line search, both as mentioned in \cite{c28}. The feedback gains computed at the final iteration is used for feedback control against noise on top of the resulting open-loop trajectory.

\noindent{\bf Comparison:} From Fig. 1 and 7, the average cost incurred for the systems in each simulation via T-PFC is close to that incurred through an NMPC approach. In other words, the cost accumulated by our perturbation linear feedback approach is nearly the same as that accumulated by an optimal deterministic control law over the feasible range of $\epsilon$ for T-PFC. T-LQR being based on the first order cost approximation, the cost rapidly diverges with increase in the noise level as reflected in Figs. 1 and 6. On the other hand, as ILQG doesn't make any claims regarding feedback, it is expected and is also clear from the same plots that the performance deteriorates rapidly with noise. 

Table I displays the time taken to execute each of the algorithms. The total execution time taken by NMPC is nearly 100 times the T-PFC in the most complex of the examples considered. The low online computational demand of T-PFC makes it scalable to implement in systems with higher dimensional state-space.

Another challenging aspect in the implementation of NMPC is generating initial guesses for online optimization. With a number of obstacle constraints or barrier functions, the NMPC optimizer fails to converge to a solution with trivial initializations and even with warm-starting, more so at higher noise levels. In contrast, T-PFC typically solves the optimization problem only once and hence, a one-time initialization is sufficient for the execution.

Unlike T-LQR, T-PFC also handles the residual second order terms of cost-to-go as well as system dynamics. This way, tuning is also bypassed as the feedback adapts itself according to the nominal cost. In contrast, TLQR can apply aggressive controls during feedback depending on LQR parameter-tuning. T-PFC in an attempt to reduce the overall cost, generates smooth and small controls relative to its nominal. This is noticeable in Fig. 3. Also, this fact plays an advantage when the nominal control is on the constraint boundary and it is undesirable for the perturbation control to deviate significantly from the nominal.

The advantage of decoupling between the optimal nominal and the perturbation feedback law is conspicuous when compared with ILQG. Parameter tuning in ILQG for regularization and line-search involves trial and error regulation and is often time consuming to searching for the right set of parameters to every given system, especially when the cost function is non-quadratic and non-polynomial. On the other hand, an NLP solver (using say, interior-point methods) can be conveniently used in a black box fashion in perturbation feedback approaches such as T-PFC (or even T-LQR) without necessitating any fine-tuning and resulting in a deterministic control policy. 

\noindent{\bf{Small noise assumption}}:
Though the theory is valid for small noise cases $i.e,$ for small epsilons, empirical results suggest a greater range of stability i.e, stability holds even for moderate levels of noise. As long as the noise falls in this range, a precise knowledge of the magnitude of noise is irrelevant as T-PFC is insensitive to noise levels.\\
{\bf Re-planning}:
At any point of time during the execution, if the cost deviates beyond a threshold from the nominal cost $i.e, C_{Th}$,  a re-planning is initiated in T-PFC. Fig. 1 (b) shows the average rate of re-planning for example-1. Until $\epsilon = 0.25$, no re-planning was necessary in the example of a car-like robot. From Fig. 1 (b), it is evident that even at above-moderate levels of noise the re-planning frequency is still eight times lesser than that required for an NMPC.\\
\noindent{\bf Limitations}:
1) T-PFC assumes a control-affine system and the cost to be in a specific form. Though many robotic systems are affine in controls, methods like T-LQR have an edge by considering a general nonlinear system. \\
2) Though T-LQR doesn't fare well on the cost incurred, it offers a flexibility to tune the feedback parameters according to ones needs, even if that means sacrificing the optimality.  

\noindent\textbf{Is NMPC necessary?} Our central observation is that the T-PFC (and even T-LQR) method has near identical performance with NMPC, while being orders of magnitude more computationally efficient, both according to the decoupling theory, as well as empirically, from the problems that we have considered here. So why not use perturbation feedback techniques instead of NMPC at least until the noise levels demand for frequent replanning?

\section{CONCLUSION}
In this paper, we have established that in a fully-observed scenario, a deterministic action policy can be split into an optimal nominal sequence and a feedback that tracks the nominal in an attempt to maintain the cost within a tube around the nominal. T-PFC maintains low cost, has low online computation and hence, faster execution. This makes our approach tractable in systems with higher dimensional states. Like MPC, the nominal trajectory design of T-PFC also allows for the inclusion of constraints as described. We have empirically shown that the overall control signals are very close to the saturation boundary, if not with-in, when the nominal is at saturation. Also, T-PFC works with minimal number of re-plannings even at medium noise levels, as against to the traditional principle of MPC to re-plan in a recurrent fashion irrespective of noise levels. Future work involves exploring the above idea of decoupling to partially-observed systems, .

\section*{APPENDIX}

{\bf Proof of Proposition 1:}
\begin{align*}
\bar{J}_t({\bf x_t}) = \mathop{min}_{{\bf u_t}} J_t({\bf x_t},{\bf u_t}) = \mathop{min}_{{\bf u_t}}\{C_t({\bf x_t}, {\bf u_t}) + \bar{J}_{t+1}({\bf x_{t+1}})\}
\end{align*}
By Taylor's expansion about the nominal state at time $t+1$,
\begin{align*}
\bar{J}_{t+1}({\bf x_{t+1}})=&\bar{J}_{t+1}({\bf \bar{x}_{t+1}}) + G_{t+1} \delta {\bf x_{t+1}} \\
&+ \frac{1}{2}\delta {\bf x_{t+1}}' P_{t+1} \delta {\bf x_{t+1}} + q_{t+1}(\delta {\bf x_{t+1}}).
\end{align*}
Substituting \(\delta {\bf x_{t+1}} = A_t \delta {\bf x_t} + B_t \delta {\bf u_t} + r_t(\delta {\bf x_t}, \delta {\bf u_t})\) in the above expansion, 
\begin{align*}
%\begin{split}
& \bar{J}_{t+1}({\bf x_{t+1}}) = \bar{J}_{t+1}({\bf \bar{x}_{t+1}}) +  G_{t+1} (A_t \delta {\bf x_t} + B_t \delta {\bf u_t} + r_t(\delta {\bf x_t}\\
&, \delta {\bf u_t}) ) + ( A_t \delta {\bf x_t} + B_t \delta {\bf u_t} + r_t(\delta {\bf x_t}, \delta {\bf u_t}))'P_{t+1} (A_t \delta {\bf x_t} \\
&+ B_t \delta {\bf u_t} +  r_t(\delta {\bf x_t}, \delta {\bf u_t}) )   + q_{t+1}(\delta {\bf x_{t+1}}).  
%\end{split}
\end{align*}
Similarly, expand the incremental cost at time $t$ about the nominal state,
\begin{align*}
C_t({\bf x_t}, {\bf u_t}) = \bar{l}_t + L_t \delta {\bf x_t} + \frac{1}{2} \delta {\bf x_t}^\intercal L_{tt} \delta {\bf x_t} + \frac{1}{2} \delta {\bf u_t}^\intercal R_t {\bf \bar{u}_t} \\
+ \frac{1}{2}  {\bf \bar{u}_t}^\intercal R_t \delta {\bf u_t} + \frac{1}{2} \delta {\bf u_t}^\intercal R_t \delta {\bf u_t}  + \frac{1}{2}{\bf \bar{u}_t}^\intercal R_t {\bf \bar{u}_t} + s_t(\delta {\bf x_t}).
\end{align*}
\begin{align*}
&J_t({\bf x_t},{\bf u_t}) = \overbrace{[\bar{l}_t + \frac{1}{2} {\bf \bar{u}_t}^\intercal R_t {\bf \bar{u}_t} + \bar{J}_{t+1}({\bf \bar{x}_{t+1}}) ]}^{\bar{J}_t({\bf \bar{x}_t})}\\
&+ \delta {\bf u_t}^\intercal(B_t' \frac{P_{t+1}}{2} B_t + \frac{1}{2} R_t) \delta {\bf u_t} + 
\delta {\bf u_t}^\intercal(B_t' \frac{P_{t+1}}{2} A_t \delta {\bf x_t} \\
&+ \frac{1}{2} R_t {\bf \bar{u}_t} +B_t' \frac{P_{t+1}}{2}r_t) + (\delta {\bf x_t}^\intercal A_t' \frac{P_{t+1}}{2}B_t 
+ \frac{1}{2} {\bf \bar{u}_t} R_t \\
&+r_t' \frac{P_{t+1}}{2}B_t + G_{t+1}B_t) \delta {\bf u_t} + \delta {\bf x_t}^\intercal A_t' \frac{P_{t+1}}{2}A_t \delta {\bf x_t} \\
&+ \delta {\bf x_t}^\intercal \frac{P_{t+1}}{2}A_t' r_t+ (r_t' \frac{P_{t+1}}{2}A_t + G_{t+1} A_t) \delta {\bf x_t} \\
&+ r_t' \frac{P_{t+1}}{2}r_t+ G_{t+1}r_t + q_t.
\end{align*}
\begin{flalign*}
\text{Now,}
\mathop{min}_{{\bf u_t}} J_t({\bf x_t}, {\bf u_t}) &=  \mathop{min}_{{\bf \bar{u}_t}}  J_t({\bf \bar{x}_t}, {\bf \bar{u}_t}) + \mathop{min}_{\delta {\bf u_t}} H_t(\delta {\bf x_t} ,\delta {\bf u_t})
\end{flalign*}
\textbf{First order optimality:} At the optimal nominal control sequence \(\bar{u}_t\), it follows from the minimum principle that
\begin{align*}
\frac{\partial C_t({\bf x_t}, {\bf u_t})}{\partial {\bf u_t}} + \frac{\partial g({\bf x_t}) }{\partial {\bf u_t}}^\intercal \frac{\partial \bar{J}_{t+1}({\bf x_{t+1}})}{\partial {\bf x_{t+1}}}  = 0   
\end{align*}
\begin{equation}
\Rightarrow R_t {\bf \bar{u}_t} + B_{t}^\intercal G_{t+1}^\intercal = 0
\end{equation}
By setting \(\frac{\partial H_t(\delta {\bf x_t}, \delta {\bf u_t})}{\partial \delta {\bf u_t}} = 0  \), we get:
\begin{align*}
 \delta {\bf u^{*}_t} &=- S_t^{-1} (R_t {\bf \bar{u}_t} + B_t^\intercal G_{t+1}^\intercal) - S_t^{-1}(B_t' P_{t+1}A_t + \\
 & (G_t \otimes \tilde R_{t,xu})^\intercal) \delta {\bf x_t}-S_t^{-1}(B_t' P_{t+1}r_t) \\
 &= \underbrace{- S_t^{-1}(B_t' P_{t+1}A_t + (G_{t+1} \otimes \tilde R_{t,xu})^\intercal)}_{K_t} \delta {\bf x_t}  \\
 &+\underbrace{S_t^{-1}(-B_t' P_{t+1}r_t)}_{p_t}
 \end{align*}
 where, \(S_t = R_t + B_t' P_{t+1}B_t.\)
 \begin{align*}
    \Rightarrow{\delta {\bf u_t}= K_t \delta {\bf x_t} + p_t}.
 \end{align*}

By substituting it in the expansion of $J_t$ and regrouping the terms based on the order of \(\delta x_t \) up to second order, we obtain:
\begin{align*}
\bar{J}_t({\bf x_t}) = \bar{J}_t({\bf \bar{x}_t}) + (L_t + (R_t {\bf \bar{u}_t} + B_t^\intercal G_{t+1}^\intercal)K_t + G_{t+1}A_t)\delta {\bf x_t} \\
+ \frac{1}{2}\delta {\bf x_t}^\intercal (L_{tt} + A_t^\intercal P_{t+1} A_t - K_t^\intercal S_t K_t + 
G_{t+1}\otimes{\tilde R}_{t,xx}) \delta {\bf x_t }.
\end{align*}

Expanding the LHS about the optimal nominal state result in the equations (2) and (3).

%following equations:
%\begin{equation}
%G_t = L_t + G_{t+1}A_t 
%\end{equation}
%\begin{equation}
%P_t = L_{tt} + A_t' P_{t+1}A_t - K_t' S_t K_t + G_{t+1}\otimes \tilde R_{t,xx}
%\end{equation}

%%%%%%%%%%%%%%%%%%%%%%%%%%%%%%%%%%%%%%%%%%%%%%%%%%%%%%%%%%%%%%%%%%%%%%%%%%%%%%%%

%\bibitem{c19} D. Tilbury, R. M. Murray, S. S. Shastry, ``Trajectory generation for the N-trailer Problem Using Goursat Normal Form," {\it IEEE Transactions on Automatic Control}, vol. 40, no. 5, May 1995.

\end{document}